\documentclass[aps,twocolumn]{revtex4}

\begin{document}

\def\bb{\begin{equation}}
\def\ee{\end{equation}}

\title{Attosecond streaking of core lines of copper dihalides}
\author{J.D. Lee}
\affiliation{School of Materials Science, Japan Advanced Institute
of Science and Technology, Ishikawa 923-1292, Japan
}
\date{\today}

\begin{abstract}
In the attosecond (as) streaking of Cu $3s$ core-level photoemission
of copper dihalides, we predict theoretically that
the satellite ($3d^9$)
is emitted later than the main line ($3d^{10}L^{-1}$; $L$: ligand).
The emission time delay is originated from the electron correlation
between the core level and $3d$ shell, which leads
to the difference in core-hole screening between satellite and main
lines. Further, we find that the time delay corresponds to
a quantification of the extrinsic loss of photoemission.
\end{abstract}
\maketitle

During the past decade, advances in attosecond (as) (1 as = $10^{-18}$ s)
laser technology have
allowed time-resolved tracking at a time scale of the order of
${\cal O}(10)$ - ${\cal O}(100)$ as of the photoinduced electron
dynamics occurring
in atoms or molecules\cite{Drescher,Swoboda,Kienberger,Krausz}.
A new perspective
has been recently opened up by attosecond streaking metrology with the
photoemission from the condensed phase\cite{Cavalieri}. Photoemission
spectroscopy is one of the most developed tools for investigating
the electronic structure of condensed matter\cite{Imada}.
The combination of
attosecond metrology and photoemission spectroscopy enables
time-domain insight into the electron photocreation and
transport stages
of the photoemission process.

In attosecond streaking photoemission, an electron emitted
by an attosecond extreme ultraviolet (XUV) pulse is streaked by the
femtosecond infrared (IR) pulse of a variable relative delay.
The first proof of the principle experiment has revealed a time delay of
about 100 as (100 $\pm$ 70 as) between photoelectrons from the localized
$4f$ core level and the conduction band of a tungsten surface\cite{Cavalieri}.
A large part of the observed time delay is found to be from
the transport delay due to a difference in group velocities of
electrons emitted from the core level and
conduction band\cite{Cavalieri,Zhang,Kazansky}.
In a subsequent attosecond photoemission experiment for the atomic
target, a different kind of time delay was observed. The slower $2s$ electrons
are found to be emitted earlier than
the faster $2p$ electrons by 21 $\pm$ 5 in neon atoms\cite{Schultze}.
This time delay is not explained by the transport, but is related
to the emission timing\cite{Schultze,Kheifets}.
The energy derivative of the quantum phase
of the dipole matrix element determines this emission time delay,
called the Wigner-Smith delay\cite{Wigner,Smith}.
The Wigner-Smith delay observed in neon atoms
could provide a new insight into intra-atomic electron correlation,
because it eventually determines
the quantum phase. This naturally tempts a challenge toward
a new time-domain understanding of electron correlation
in a true solid with strongly correlated electrons.

\begin{figure}
\vspace*{5.5cm}
\includegraphics{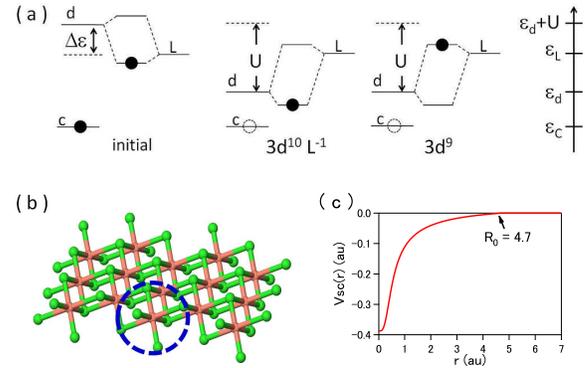}
\caption{(a) Copper dihalides before and after photoemission.
For instance, $d$ implies a Cu $3d$ state, $c$ a Cu ($3s$) core level,
and $L$ a ligand, i.e., $L=(3p)^6$ of Cl for CuCl$_2$.
$U$ is the electron correlation between $c$ and $d$.
(b) Crystal structure of CuCl$_2$:
a CuCl$_6$ cluster is designated by a blue dashed circle.
(c) $V_{\rm sc}(r)$ for CuCl$_2$.
}
\label{FIG1}
\end{figure}

Copper dihalides are insulating solids of $3d$ transition metal compounds,
the simplest and best understood
systems where the charge transfer scenario is applied\cite{Laan}.
Figure \ref{FIG1}(a) shows that photoexcitation
of the Cu core level results in two final states.
The $3d^9$ final state leaves the valence configuration largely unaltered
and leads to a satellite (excited one), whereas
the $3d^{10}L^{-1}$ final state ($L$: ligand shell)
transfers an electron from a ligand to
a metal ion and leads to the
main line (ground one)\cite{Huefner}.
Within this simple model,
the extrinsic loss of photoelectron, that is, {\it loss effects
due to an interaction of photoelectron with the remaining solid
during its travel to the surface}, was studied in copper
dihalides by Lee {\it et al.}\cite{Lee}. Their work could
provide the key inspiration to the problem of attosecond streaking
photoemission of strongly correlated electron systems.

In this Letter,
we study the attosecond streaking of $3s$ core lines of copper dihalides
by explicitly considering the photoelectron scattering
due to the core hole left after photoemission.
We find that electron correlation between the core level and
$3d$ shell leads to a difference in core-hole screening
between the satellite and the main lines, which enables
the theoretical prediction that
the satellite is emitted later than the main line, say by 90 as (for CuBr$_2$),
73 as (for CuCl$_2$), and 40 as (for CuF$_2$) at $\omega_{\rm XUV}=141$ eV,
where $\omega_{\rm XUV}$ is the XUV photon energy.
The emission time delay strongly depends on the XUV
photon energy and disappears at just $\sim 20$ eV over the threshold
(i.e., the binding energy of the satellite).
We further find that the observed time delay is a quantification of
the extrinsic loss of photoemission.

\begin{figure}
\vspace*{7.1cm}
\includegraphics{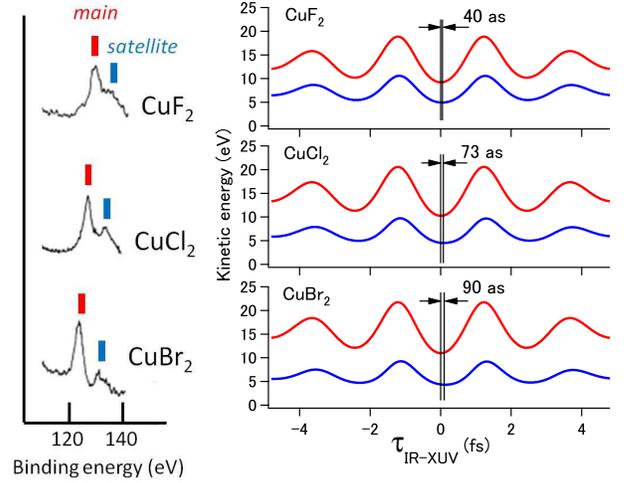}
\caption{Left: Photoemission spectra of Cu 3s core lines of
copper dihalides\cite{Laan}. Right: Centers of energy (COEs)
of streaked photoemission spectra
of copper dihalides with respect to a relative delay $\tau_{\rm IR-XUV}$
of the IR pulse from the XUV pulse ($\omega_{\rm XUV}=141$ eV).
Main lines (red lines) are prior to
the satellites (blue lines).
}
\label{FIG2}
\end{figure}

Photoemission analysis of strongly correlated electron systems
including $d$ electrons could be often simplified from
their localized nature.
The simplest, but quite successful model based on
a molecular orbital approach, represents
the $d$ part of the system by one orbital, and the ligand
by another\cite{Huefner}.
We start from a Hamiltonian ${\cal H}_0$ of three-state
(two-state + one core level) model
describing copper dihalides, being  consistently with Fig.\ref{FIG1}(a),
\bb\label{Eq:1}
{\cal H}_0=\varepsilon_d n_d+\varepsilon_L n_L+\varepsilon_C n_C+Un_Cn_d+
t(c_d^{\dagger}c_L+c_L^{\dagger}c_d),
\ee
where the first two terms depict energy levels of the Cu $3d$ shell
($n_d=c_d^{\dagger}c_d$)
and a halide ligand $L$ ($n_L=c_L^{\dagger}c_L$)
and the last term depicts
the hybridization between them. The remaining terms involve
the energy level of Cu $3s$ core level ($n_C=c_C^{\dagger}c_C$)
and the electron correlation $U$
between the core level and the $3d$ shell. From Eq.(1), the initial
ground state $|\psi_0\rangle$ is readily obtained by inserting $n_C=1$ as
$|\psi_0\rangle=-\sin\theta|C\rangle|d\rangle+\cos\theta|C\rangle|L\rangle$
with $\tan2\theta=2t/(\varepsilon_d+U-\varepsilon_L)$.
The ground state energy $E_0$ is
$E_0=\varepsilon_C+\frac{1}{2}(\varepsilon_d+U+\varepsilon_L)
-\frac{1}{2}\sqrt{(\varepsilon_d+U-\varepsilon_L)^2+4t^2}$.
The final states of the target (after photoemission)
are also immediately obtained by inserting $n_C=0$ as
$|\psi_{\rm m}\rangle=\cos\varphi|d\rangle-\sin\varphi|L\rangle$
and $|\psi_{\rm s}\rangle=\sin\varphi|d\rangle+\cos\varphi|L\rangle$
with $\tan2\varphi=2t/(\varepsilon_L-\varepsilon_d)$, corresponding
to the main ($3d^{10}L^{-1}$) and satellite state ($3d^9$), respectively,
as illustrated in Fig.\ref{FIG1}(a).
The energy eigenvalues $E_{\rm m}$
and $E_{\rm s}$ are $E_{\rm s}^{\rm m}=\frac{1}{2}(\varepsilon_d+\varepsilon_L)
\mp\frac{1}{2}\delta E$, where $\delta E$ is the separation between
two states given by
$\delta E=\sqrt{(\varepsilon_d-\varepsilon_L)^2+4t^2}$.
Then photocurrent is
$J(\omega)=|w_{\rm m}|^2\delta(\omega+E_0-E_{\rm m})
+|w_{\rm s}|^2\delta(\omega+E_0-E_{\rm s})$, where
$w_{\rm m}=-\sin(\varphi+\theta)$ and $w_{\rm s}=\cos(\varphi+\theta)$
from $w_{\rm m(s)}=\langle \psi_{\rm m(s)}|c_C|\psi_0\rangle$.
$J(\omega)$ can be directly compared with the
left panel of Fig.\ref{FIG2} so
that the material parameters would be taken from
the relative peak strengths and peak positions
in the experiment\cite{parameters}.
This analysis constructs the sudden approximation.

${\cal H}_1$ comprises the excitation of photoelectron
by the XUV pulse and its streaking
by the IR pulse,
\begin{eqnarray}\label{Eq:2}
{\cal H}_1&=&\sum_{\bf k}\left[\varepsilon_{\bf k}
           -{\bf k}\cdot{\bf A}_{\rm IR}(\tau+\tau_{\rm IR-XUV})\right]n_{\bf k}
\nonumber \\
          &+&\sum_{\bf k}m_{\bf k}
           (c_{\bf k}^{\dagger}c_C+c_C^{\dagger}c_{\bf k})A_{\rm XUV}(\tau),
\end{eqnarray}
where $n_{\bf k}=c_{\bf k}^{\dagger}c_{\bf k}$ is the occupation number of
photoelectron with kinetic energy of $\varepsilon_{\bf k}={\bf k}^2/2$.
${\bf A}_{\rm IR}(\tau)$ is the streaking IR pulse given by
${\bf A}_{\rm IR}(\tau)=0.2\exp[-3.24\times 10^{-5}\tau^2]
\cos(0.0608\tau)\hat{\bf e}_{\rm IR}$
and ${\bf A}_{\rm XUV}(\tau)$ is the photoexciting XUV pulse given by
${\bf A}_{\rm XUV}(\tau)=A_{\rm XUV}^0\exp[-0.036\tau^2]
\cos(\omega_{\rm XUV}\tau)\hat{\bf e}_{\rm XUV}$.
XUV and IR pulses are assumed to be linearly polarized along the direction
of photoelectron detection.
$\bar{\tau}_{\rm IR}$ and $\bar{\tau}_{\rm XUV}$ are the half width
at half maximum (HWHMs) of the IR and XUV pulses, i.e.,
$\bar{\tau}_{\rm IR}=146.26 \ {\rm a.u.}=3.5$ fs and
$\bar{\tau}_{\rm XUV}=4.39 \ {\rm a.u.}=105$ as (where a.u. means atomic unit).
$\tau_{\rm IR-XUV}$ is the relative delay of the IR pulse from the XUV pulse.
$m_{\bf k}$ is $\langle {\bf k}|\Delta|C\rangle$ of the dipole
operator $\Delta$ ($={\bf r}\cdot \hat{\bf e}_{\rm XUV}$)
with $|{\bf k}\rangle$ as the plane wave for the photoelectron,
which is equivalent to the so-called strong field approximation.
Noting that the ${\bf k}$-dependence of $m_{\bf k}$ is usually weak, 
we take $m_{\bf k}$ just as a constant.

Beyond the sudden approximation, we should consider the photoelectron
scattering ${\cal V}$ by the core hole as\cite{Lee}
\begin{eqnarray}\label{Eq:3}
{\cal V}&=&\sum_{{\bf k}{\bf k}^{\prime}}
         \left[n_dV^{(d)}_{{\bf k}{\bf k}^{\prime}}
              +n_LV^{(L)}_{{\bf k}{\bf k}^{\prime}}
              -V^{(C)}_{{\bf k}{\bf k}^{\prime}}\right]
              c_{\bf k}^{\dagger}c_{{\bf k}^{\prime}}
\nonumber \\
        &\approx&n_L\sum_{{\bf k}{\bf k}^{\prime}}
        V_{{\bf k}{\bf k}^{\prime}}^{\rm sc}
        c_{\bf k}^{\dagger}c_{{\bf k}^{\prime}},
\end{eqnarray}
where $V_{{\bf k}{\bf k}^{\prime}}^{(\nu)}(=\int d{\bf r}
\psi_{\bf k}^{\ast}({\bf r})
V^{(\nu)}(r)\psi_{{\bf k}^{\prime}}({\bf r}))$ is the matrix element of the
Coulomb potential from the orbital $\nu$ using the plane wave basis.
The second line is obtained by
assuming that the Cu $3d$ level is so local as to have
$V^{(C)}(r)\approx V^{(d)}(r)$ and noting $n_d+n_L=1$. We now have
$V_{{\bf k}{\bf k}^{\prime}}^{\rm sc}=\int d{\bf r}\psi_{\bf k}^{\ast}({\bf r})
V_{\rm sc}(r)\psi_{{\bf k}^{\prime}}({\bf r})$ and
$V_{\rm sc}(r)=V^{(L)}(r)-V^{(d)}(r)$. $V^{(d)}(r)$ for the Cu $3d$ orbital,
which can be calculated by generating the charge densities in terms of
Slater's orbitals\cite{Slater}, while $V^{(L)}(r)$ would be approximated
by a spherical charged shell made of six ligand orbitals
of a CuCl$_6$ cluster (for CuCl$_2$) with the radius $R_0$
as in Fig.\ref{FIG1}(b). $R_0$ is an average distance between Cu and Cl.
In the lowest approximation, there would be no scattering outside the cluster
before reaching the surface unlike the metallic case\cite{cluster}.
This gives $V_{\rm sc}(r)=\frac{1}{\epsilon}
\left[-V^{(d)}(r)+\frac{1}{R_0}\right]\Theta(R_0-r)$,
where $\Theta(R_0-r)$
is the Heaviside step function\cite{parameters}.
$V_{\rm sc}(r)$ for CuCl$_2$ is displayed
in Fig.\ref{FIG1}(c).

Under the total Hamiltonian of ${\cal H}_0+{\cal H}_1+{\cal V}$,
by solving the time-dependent Schr\"{o}dinger equation with
$A_{\rm XUV}^0\to 0$, we can treat
the streaking of photoemission in an exact fashion. The total wave function
$|\Psi(\tau)\rangle$ at time $\tau$ can be written as
\begin{eqnarray}\label{Eq:4}
|\Psi(\tau)\rangle&=&C_d(\tau)|d\rangle|C\rangle+C_L(\tau)|L\rangle|C\rangle
\nonumber \\
                  &+&\sum_{\bf k}C_{d{\bf k}}|d\rangle|{\bf k}\rangle
                  +\sum_{\bf k}C_{L{\bf k}}|L\rangle|{\bf k}\rangle.
\end{eqnarray}
Dynamics will start by turning on the XUV pulse so that the initial state
should be $|\Psi(\tau_0)\rangle=|\psi_0\rangle$ given at $\tau=\tau_0\ll
-\bar{\tau}_{\rm XUV}$. The streaked photoemission spectrogram
could be calculated for the main and satellite lines as follows:
$S_{\rm m}({\bf k},\tau_{\rm IR-XUV})=|\langle{\bf k}|\langle\psi_{\rm m}
                  |\Psi(\tau_{\rm max})\rangle|^2$
and
$S_{\rm s}({\bf k},\tau_{\rm IR-XUV})=|\langle{\bf k}|\langle\psi_{\rm s}
                  |\Psi(\tau_{\rm max})\rangle|^2$
for a given value of $\tau_{\rm IR-XUV}$.
Spectrograms are found to converge
at $\tau=\tau_{\rm max}\gg \bar{\tau}_{\rm XUV}$. In the actual calculation,
we adopt $\tau_0=-15$ a.u. and $\tau_{\rm max}=100$ a.u..

\begin{figure}
\vspace*{5.8cm}
\includegraphics{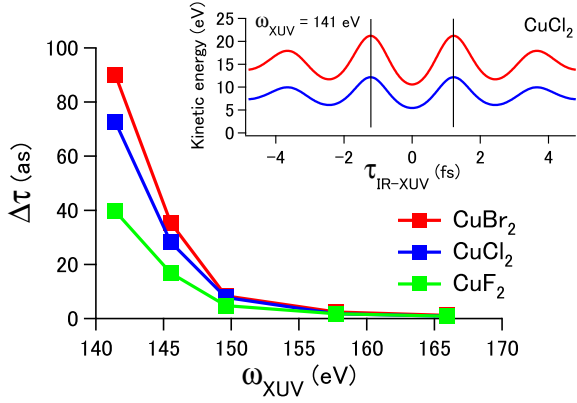}
\caption{Emission time delays $\Delta\tau$ between satellite and main lines
for CuBr$_2$, CuCl$_2$, and CuF$_2$ with respect to $\omega_{\rm XUV}$.
In the inset, COEs of streaked photoemission spectra
within the sudden approximation,
i.e., ${\cal V}=0$, are given at $\omega_{\rm XUV}=141$ eV for CuCl$_2$.
}
\label{FIG3}
\end{figure}

In the streaked spectrogram $S_{\rm m}({\bf k},\tau_{\rm IR-XUV})$
and $S_{\rm s}({\bf k},\tau_{\rm IR-XUV})$ with distributions of
photoelectron kinetic energy, we determine the center of energy (COE).
COEs of streaked photoemission spectra are presented
with respect to $\tau_{\rm IR-XUV}$
at $\omega_{\rm XUV}=141$ eV in the right panel of Fig.\ref{FIG2}.
The metrology of streaked spectra reveals a time delay
$\Delta\tau$ in the emission of electrons corresponding to the satellite
with respect to those corresponding to the main line.
This finding is dramatic. Main and satellite lines are not
simultaneously created, but satellites are emitted later than main lines
by $\Delta\tau=40$ as, 73 as, and 90 as at $\omega_{\rm XUV}=141$ eV
for CuF$_2$, CuCl$_2$, and CuBr$_2$, respectively.

The emission time delay $\Delta\tau$ strongly depends
on the XUV pulse energy
$\omega_{\rm XUV}$ as shown in Fig.\ref{FIG3}.
Notably, $\Delta\tau$ already disappears
at $\omega_{\rm XUV}-E_{\rm s}^B\gtrsim 20$ eV,
where $E_{\rm s}^B$ is the binding energy of the satellite,
$E_{\rm s}^B=E_{\rm s}-E_0\approx 131$ eV, which is almost the same
for the three compounds, as shown in the left panel of Fig.\ref{FIG2}.
The same scattering potential ${\cal V}$ also
gives rise to extrinsic loss of photoelectron
in photoemission
spectroscopy. Extrinsic loss would give a change in
relative peak strengths or an asymmetric broadening of peaks
observed in the spectroscopy.
It was argued that extrinsic loss due to
extended excitations in a metal sustains up to the photon energy
$\sim {\cal O}(1)$ keV\cite{Hedin}, whereas that due to local
excitations in an electron-correlated insulator
disappears at around $10-20$ eV above the
threshold\cite{Lee}. In the inset of Fig.\ref{FIG3},
COEs of streaked photoemission spectra within the sudden approximation, i.e.,
putting ${\cal V}=0$, at $\omega_{\rm XUV}=141$ eV
for CuCl$_2$ are given, so that
$\Delta\tau=0$ is in fact confirmed.
They indicate that
$\Delta\tau$ could be a quantification of extrinsic
loss contribution to the photoemission spectra of a given system.

\begin{figure}
\vspace*{4.cm}
\includegraphics{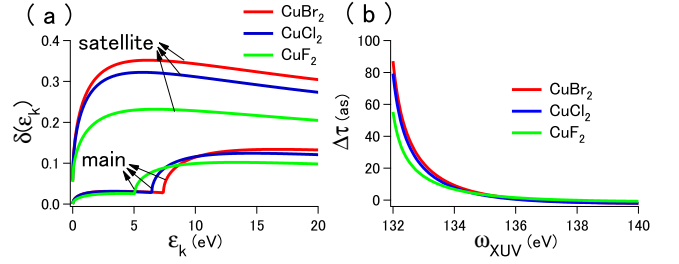}
\caption{(a) Energy-dependent phases of dipole matrix elements
corresponding to the satellite and main lines
of CuBr$_2$, CuCl$_2$, and CuF$_2$. (b) Emission time delays $\Delta\tau$
between satellite and main lines.
Photoelectron scattering is
taken into account within the first-order perturbation theory.
}
\label{FIG4}
\end{figure}

More figurative understanding of $\Delta\tau$ is possible
by noting that the Wigner-Smith delay is determined by
the energy derivative of the quantum phase of the matrix element.
The exact final state $|\Psi_{\rm m(s)}^{\bf k}\rangle$
in the scattering theory is
$|\Psi_{\rm m(s)}^{\bf k}\rangle=\left[1+
\frac{1}{E-{\cal H}_0-{\cal T}-{\cal V}-i0^+}{\cal V}\right]
|\psi_{\rm m(s)}\rangle|{\bf k}\rangle$, where
$E$ is the energy of the final state, given by
$E=\varepsilon_{\bf k}+E_{\rm m(s)}$ for the main and satellite,
respectively, and
${\cal T}=\sum_{\bf k}\varepsilon_{\bf k}c_{\bf k}^{\dagger}
c_{\bf k}$. To the first order in ${\cal V}$, we neglect ${\cal V}$
in the denominator. We can then obtain the matrix element
$M_{\rm m(s)}({\bf k})=\langle \Psi_{\rm m(s)}^{\bf k}|
\Delta|\psi_0\rangle$.
The quantum phase of the matrix element will be
$\delta_{\rm m(s)}(\varepsilon_{\bf k})=\arg[M_{\rm m(s)}
({\bf k})]$, the behavior of which
is given in Fig.\ref{FIG4}(a). The time delay $\Delta\tau$ between
the satellite and main lines is calculated by $\Delta\tau=\partial
\delta_{\rm s}(\varepsilon_{\bf k})/\partial\varepsilon_{\bf k}-
\partial\delta_{\rm m}(\varepsilon_{\bf k})\partial\varepsilon_{\bf k}
|_{\varepsilon_{\bf k}+\delta E}$ and given with respect to
$\omega_{\rm XUV}$ in Fig.\ref{FIG4}(b), which could be compared
with Fig.\ref{FIG3}. Suppression of $\Delta\tau$ with $\omega_{\rm XUV}$
is much faster in the first-order perturbation so that
$\Delta\tau$ disappears at $\omega_{\rm XUV}-E_{\rm s}^B\gtrsim 5$ eV.
Perturbation theory cannot give quantitatively good results
compared to the exact ones in Figs.\ref{FIG2} and \ref{FIG3}
due to the strong scattering potential,
but could give intuitive understanding that
the Wigner-Smith delay occurs mostly in satellite, especially
in the low energy region, i.e.,
$\partial\delta_{\rm s}(\varepsilon_{\bf k})/\partial\varepsilon_{\bf k}
\gg \partial\delta_{\rm m}(\varepsilon_{\bf k})\partial\varepsilon_{\bf k}
|_{\varepsilon_{\bf k}+\delta E}$.

\begin{figure}
\vspace*{4.6cm}
\includegraphics{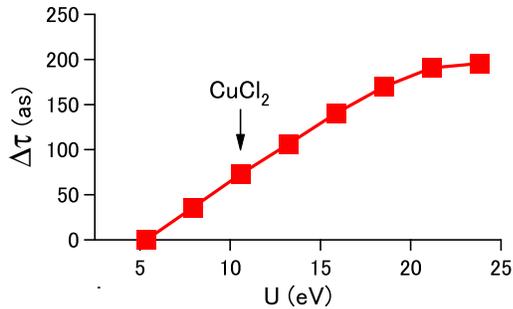}
\caption{Emission time delays $\Delta\tau$ with respect to $U$
at $\omega_{\rm XUV}=141$ eV obtained by
keeping the same initial state by fixing the value of
$\varepsilon_d+U$ ($=2.4$ eV as in CuCl$_2$).
All other material parameters are taken from CuCl$_2$.
}
\label{FIG5}
\end{figure}

A strong electron correlation between the core level and the $3d$ shell plays
an essential role in inducing the emission time delay between the main
and satellite lines in copper dihalides.
The electron correlation $U$ leads to two final states with different
screening of the core hole created after photoemission.
The larger the $U$, the greater the difference.
When keeping the same initial state by fixing the value of $\varepsilon_d+U$,
as illustrated in Fig.\ref{FIG1}(a),
we may have $|\psi_{\rm m}\rangle\to
|d\rangle$ and $|\psi_{\rm s}\rangle\to |L\rangle$
in a limit of large $U\to \infty$, and
$|{\bf k}\rangle|\psi_{\rm m}\rangle $ will not be scattered
by the scattering potential ${\cal V}$, i.e., {\it perfect screening},
while $|{\bf k}\rangle|\psi_{\rm s}\rangle$
is maximally scattered, i.e., {\it no screening}.
This will cause an increase of $\Delta\tau$, which
is clearly demonstrated in Fig.\ref{FIG5}.
It should also be noted that the correlation $U$ scales the potential strength,
i.e., $U=-V_{\rm sc}(0)$\cite{parameters}.
From $\Delta\varepsilon \approx U$
under an assumption of small hybridization (See Fig.\ref{FIG1}(a)),
main and satellite lines would overlap at $U\approx 5.3$ eV
and $\Delta\tau$ would then vanish.

Finally, we note that the transport delay occurring inside the solid
was not considered here. However, such transport delay could be
separated from the Wigner-Smith delay because it stems simply from
different group velocities between satellite and main lines
without additional scattering\cite{cluster}. Further,
if an atom-thick adsorbate is technically possible with copper dihalides,
the transport delay may in principle be removed in the experiment.

To summarize, we explored the attosecond streaking of $3s$ core lines of copper
dihalides by incorporating a model of copper dihalide cluster to
describe the photoelectron scattering.
We found that in the emission timing, the satellite comes
later than the main line. Such an emission time delay
originates from the electron correlation between the core level
and the $3d$ shell, which leads to different
core-hole screening depending on satellite and main lines.
The time delay strongly depends on
$\omega_{\rm XUV}$ and its disappearance is energetically rapid, i.e.,
at just $\sim 20$ eV above the threshold. To our best knowledge,
this is the first study to explore the emission timing of electrons
in a true solid with strong electron correlation.
Lastly, we claim that
the time delay has the same root in the extrinsic loss of photoelectrons
and it is a quantification of extrinsic loss.

We thank Vladislav Yakovlev for useful discussions and
critical reading of the manuscript.
This work was supported by KAKENHI-23540364
from MEXT, Japan.


\begin{references}
\bibitem{parameters} The necessary parameters for copper dihalides are
                     well tabulated in Ref.\cite{Lee}: $U=8.62$ eV,
                     $t=1.63$ eV, $\delta E=5.41$ eV,
                     $\varepsilon_L=-4.3$ eV, and $R_0=3.86$ a.u.
                     for CuF$_2$, $U=10.58$ eV,
                     $t=1.84$ eV, $\delta E=6.45$ eV,
                     $\varepsilon_L=-2.9$ eV, and $R_0=4.71$ a.u.
                     for CuCl$_2$, $U=12.33$ eV,
                     $t=2.02$ eV, $\delta E=7.37$ eV,
                     $\varepsilon_L=-1.96$ eV, and $R_0=5.01$ a.u.
                     for CuBr$_2$, respectively. The value
                     $\varepsilon_C=-130.6$ eV
                     is taken as common to three compounds.
                     Finally, $\epsilon$ is
                     a constant to match $V_{\rm sc}(0)=-U$:
                     $\epsilon=2.26$, 1.96, and 1.71 are found for
                     CuF$_2$, CuCl$_2$, and CuBr$_2$, respectively.
\bibitem{cluster} In contrast, in a metal,
                  scattering would occur via periodic ionic potential,
                  channels of electron-hole pairs and plasmon,
                  etc., before escaping through the surface.
\end{references}
\end{document}